\documentclass[twocolumn,conference]{IEEEtran}
\usepackage[numbers,sort&compress]{natbib}
\usepackage{graphicx,xcolor}
\usepackage{amsfonts,amssymb,amstext,amsthm,mathtools}
\usepackage{enumerate,enumitem}
\usepackage{algorithm}
\usepackage{algpseudocode}
\usepackage{dsfont,amsthm}
\usepackage{amscd}
\usepackage{bm}
\usepackage{color}
\usepackage{float} % This package allows the H specifier
\usepackage{makecell}
\usepackage[acronym]{glossaries}
\newacronym{dac}{DAC}{digital-to-analog converter}
\newacronym{pam}{PAM}{pulse amplitude modulation}
\newacronym{mimo}{MIMO}{multiple-input multiple-output}
\newacronym{rf}{RF}{radio frequency}
\newacronym{pa}{PA}{power amplifier}
\newacronym{papr}{PAPR}{peak-to-average power ratio}
\newacronym{awgn}{AWGN}{additive white Gaussian noise}
\newacronym{psd}{PSD}{power spectral density}
\newacronym{pts}{PTS}{probabilistic temporal shaping}

\newtheorem{theorem}{Theorem}

\newcommand{\E}[1]{{\mathbb{E}}\left[#1\right]}

\newcommand{\sinc}{\operatorname{sinc}}
\newcommand{\si}{\operatorname{Si}}

\newcommand{\sgn}{\operatorname{sgn}}

\newcommand{\adj}{\operatorname{adj}}

\newcommand{\SNR}{{\sf SNR}}

\long\def\comment#1{}

\newcommand{\av}{{\bf a}}
\newcommand{\bv}{{\bf b}}
\newcommand{\cv}{{\bf c}}
\newcommand{\dv}{{\bf d}}

\newcommand{\uv}{{\bf u}}

\newcommand{\xv}{{\bf x}}

\newcommand{\zv}{{\bf z}}

\newcommand{\Av}{{\bf A}}

\newcommand{\Cv}{{\bf C}}

\newcommand{\Am}{{\bf A}}

\newcommand{\Cm}{{\bf C}}
\newcommand{\Dm}{{\bf D}}

\newcommand{\Id}{{\bf I}}
\newcommand{\Jm}{{\bf J}}

\newcommand{\Mm}{{\bf M}}

\newcommand{\Sm}{{\bf S}}

\newcommand{\Vm}{{\bf V}}
\newcommand{\Xm}{{\bf X}}
\newcommand{\Ym}{{\bf Y}}
\newcommand{\Zm}{{\bf Z}}

\newcommand{\omegav}{\hbox{\boldmath$\omega$}}

\newcommand{\Omegam}{\hbox{\boldmath$\Omega$}}

\usepackage[
    bookmarks=true,         % show bookmarks bar?
    unicode=false,          % non-Latin characters in AcrobatÕs bookmarks
    pdftoolbar=true,        % show AcrobatÕs toolbar?
    pdfmenubar=true,        % show AcrobatÕs menu?
    pdffitwindow=false,     % window fit to page when opened
    pdfstartview={FitH},    % fits the width of the page to the window
    pdfnewwindow=true,      % links in new window
    colorlinks=true,       % false: boxed links; true: colored links
    linkcolor=red,          % color of internal links (change box color with linkbordercolor)
    citecolor=green,        % color of links to bibliography
    filecolor=magenta,      % color of file links
    urlcolor=cyan           % color of external links
]{hyperref}
\begin{document}

\title{Probabilistic Temporal Shaping for Level-Constrained Signaling on Bandlimited Additive White Gaussian Noise Channels}

\author{%
    \IEEEauthorblockN{Mahdi~Mahvari\IEEEauthorrefmark{1}, Gerhard~Kramer\IEEEauthorrefmark{1}, and Shlomo Shamai (Shitz)\IEEEauthorrefmark{2}}
    \IEEEauthorblockA{\IEEEauthorrefmark{1}%
    School of Computation, Information and Technology, Technical University of Munich, 80333 Munich, Germany}
   \IEEEauthorblockA{\IEEEauthorrefmark{2}%
    Dept. of Electrical and Computer Engineering, Technion—Israel Institute of Technology, Haifa 3200003, Israel}
}

%\thanks{
%Date of current version \today.
%This work was supported by the German Research Foundation (DFG) through project 509917421.
%This paper was presented in part at the 20?? IEEE Symposium on ... [DOI: ].
%}
%\thanks{
%M. Mahvari and G. Kramer are with the Institute for Communications Engineering, School of Computation, Information and Technology, Technical University of Munich (TUM), 80333 Munich, Germany (e-mail: mahdi.mahvari@tum.de; gerhard.kramer@tum.de).
%}
%}

\maketitle
\thispagestyle{plain}
\pagestyle{plain}

\begin{abstract}
    Level-constraints model one-bit-quantized signaling over real-alphabet continuous-time channels. New lower bounds on the capacity of bandlimited, additive white Gaussian noise channels with level-constrained inputs are derived by using probabilistic temporal shaping (PTS). The optimal shaping density is derived for signals with one data-dependent sign change per Nyquist-rate sample, which ensures they satisfy an invertibility requirement. Calculations show that PTS improves the best existing lower bound by at least 1.94 dB at high signal-to-noise ratio (SNR). A simpler sequential PTS scheme achieves a gain of 1.64 dB at high SNR.
\end{abstract}


%%%%%%%%%%%%%%%%%%%%%%%%%%%%%%%%%%% %%%%%%%%%%%%%%%%%%%%%%%%%%%%%%%%%%%
\section{Introduction}
\label{sec:introduction}

\Glspl{dac} and \glspl{pa} are power-intensive devices. The power consumed by \glspl{dac} scales linearly with the sampling rate and exponentially with the resolution \cite{walden1999performance, sundstrom2008power}. The latter scaling can be a limiting factor, e.g., in \gls{mimo} transmission with many \gls{rf} chains \cite{rusek2012scaling,lu2014overview}. A \gls{pa} also limits performance by introducing nonlinear distortions for high-\gls{papr} signals \cite{cripps2006rf}.

Two remedies are hybrid digital-analog structures and oversampled low-resolution quantizers. The hybrid approach reduces the number of \glspl{dac}, and is useful when the channel matrix is low-rank or ill-conditioned. Oversampling, instead, allows reducing the \gls{dac} resolution and the \gls{papr}. For example, one may use sigma-delta ($\Sigma\Delta$) modulators with high sampling rates and peak-constrained signaling.

This paper studies bandlimited channels with \gls{awgn} and level constraints. The paper is organized as follows. Sec.~\ref{sec:problem} formulates the problem and Sec.~\ref{sec:coding-schemes} reviews coding schemes and capacity bounds. Sec.~\ref{sec:main} develops our main results. We first show that amplitude modulation does not help for schemes with one data-dependent sign change per Nyquist-rate sample. We then introduce \gls{pts}, determine the optimal shaping density, and prove a theorem on the entropy of the noise-free process at the output of the bandlimited channel. Calculations show that \gls{pts} improves the best existing lower bound on capacity by at least $1.94$ dB at high signal-to-noise ratio ($\SNR$). We also introduce a sequential \gls{pts} scheme that gains $1.64$ dB at high $\SNR$. Sec.~\ref{sec:conclusion} concludes the paper.

%%%%%%%%%%%%%%%%%%%%%%%%%%%%%%%%%%%
%%%%%%%%%%%%%%%%%%%%%%%%%%%%%%%%%%%
\section{Problem Formulation}
\label{sec:problem}

\begin{figure}[t]
    \centering
    \includegraphics[width = 0.45\textwidth]{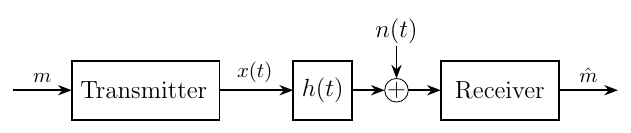}
    % \scalebox{0.7}{\input{figures/}}
    \caption{AWGN channel with a bandlimited filter.}
    \label{fig:awgn_model}
\end{figure}

Fig. \ref{fig:awgn_model} shows the model. The transmitter maps a message $m$ to a continuous-time signal $x(t)$ that is filtered by a channel response $h(t)$ with spectrum $H(f)$. We study an ideal low-pass filter with bandwidth $W$:
\begin{align}
    h(t) & = 2W \sinc(2Wt) = \sin(2\pi W t)/(\pi t)  \label{eq:ideal-LPF-t} \\
    H(f) & = 1(|f| \le W)
    \label{eq:ideal-LPF}
\end{align}
where $1(.)$ is the indicator function. The filter output is perturbed by bandlimited noise $n(t) = w(t)*h(t)$, where $w(t)$ is \gls{awgn} with two-sided \gls{psd} $N_0/2$ Watts/Hz.

Under the average power constraint
\begin{align}
    \lim_{T\rightarrow \infty} \frac{1}{T} \int_0^T x^2(t) \, dt \le P_\text{avg}
    \label{eq:avgP}
\end{align}
the capacity $C$ is achieved by choosing $x(t)$ to be a realization of a Gaussian process with bandwidth $W$. In practice, there is also a peak constraint $|x(t)| \le \sqrt{P}$. The level constraint $x(t) = \pm \sqrt{P}$ is even more restrictive, and has been studied for magnetic storage \cite{ozarow1988achievable, heegard1992bounding, chayat1999bounds}, where $ h(t)$ is a differentiator followed by a low-pass filter \cite{ozarow1988achievable}. 

The bandlimited capacity under peak or level constraints seems difficult to compute. Remarkably, the paper \cite{ozarow1988achievable} shows that the level-constrained capacity $C_L$ is the same as the peak-constrained capacity for channel responses with finite energy, i.e., $\|h(t)\|^2<\infty$. Recall that the capacity of the channel \eqref{eq:ideal-LPF} under the constraint \eqref{eq:avgP} is
\begin{align*}
    C/W = \log\left(1+\SNR\right)
    %\label{eq:avg_capacity}
\end{align*}
where $\SNR=P/(N_0W)$. We study bounds of the form
\begin{align*}
    C_L/W \gtrless
    \log\left( 1 + \gamma\, \SNR\right)
    %\label{eq:gamma_capacity}
\end{align*}
where the power factor $\gamma$ satisfies $0<\gamma<1$. The following bounds appear in \cite{ozarow1988achievable,shamai1989upper,peleg2021information,tzachy2025upper}. 

\begin{theorem}
    \label{thm:bounds}
    For the ideal low-pass filter $h(t)$ in \eqref{eq:ideal-LPF} we have
    \begin{align*}
        \log(1+ \gamma_1\, \SNR)
        \le C_L/W
        \le \log(1+ \gamma_2\, \SNR) % 0.934  \label{eq:shamai1989upper}
        %\label{eq:CL-bounds} 
    \end{align*}
    where $\gamma_1=0.2586$ and $\gamma_2=0.9259$.
\end{theorem}

%%%%%%%%%%%%%%%%%%%%%%%%%%%%%%%%%%% %%%%%%%%%%%%%%%%%%%%%%%%%%%%%%%%%%%
\section{Coding Schemes}
\label{sec:coding-schemes}

We review the lower bounds on capacity in \cite{peleg2021information}, which are based on signals $x(t)$ that have at most one data-dependent sign change per Nyquist-rate sampling interval $T=1/(2W)$. The coding schemes permit infinite precision for the switching times; see also~\cite{peleg2024geometrical,dabora2024achievable,tzachy2025upper}.

The first bounding step decouples the signal and noise using the entropy power inequality (EPI). Let $\Ym$ and $\Zm$ be the Nyquist-rate sampled outputs of the channel with and without noise, respectively. The EPI gives
\begin{align*}
    h(\Ym) \ge \frac{N}{2} \log\left(e^{2h(\Zm)/N} + 2\pi e \sigma_n^2\right)
    %\label{eq:EPI}
\end{align*}
where $N$ is the block length and $\sigma_n^2=N_0 W$. We thus have
\begin{align}
    \frac{1}{N}I(\Ym;\Xm)&=\frac{1}{N}h(\Ym) - \frac{1}{N}h(\Ym|\Xm) \nonumber\\
    &\ge \frac{1}{2} \log\left(e^{2h(\Zm)/N} + 2\pi e \sigma_n^2\right) - \frac{1}{2}\log(2 \pi e \sigma_n^2)\nonumber\\
    & = \log(1+\gamma_1\; \SNR)
    \label{eq:EPI2}
\end{align}
where $\gamma_1 = e^{2h(\Zm)/N}/(2\pi e)$.

The authors of \cite{peleg2021information} next design increasingly sophisticated schemes, labeled A-D, to increase $h(\Zm)$. To describe them, suppose there are $N$ data-dependent sign changes at times 
$$\cv = (c_1,c_2,\dots,c_N)$$
and $N_d$ deterministic sign changes at times 
$$\dv = (d_1,d_2,\dots,d_{N_d})$$
where the entries of $\cv$ and $\dv$ are strictly increasing in the time interval $[0,NT)$. Let $\cv'$ be the ordering of $(\cv, \dv)$, so that $0\le c_1' < c_2' < \dots < c_{N+N_d}'< NT$. The level-constrained signals we study have the form
\begin{align}
     x(t) & = u(t) + \sum_{i=1}^{N+N_d}
     (-1)^{i}\, 2\, u(t-c_i') \nonumber \\
     & \quad + (-1)^{N+N_d+1}\, u(t-NT)
     \label{eq:schemes}
\end{align}
where $u(t)$ is the Heaviside step function so that $|x(t)|\le 1$. The output of the bandlimited filter \eqref{eq:ideal-LPF-t} is
\begin{align}
    z(t) = \int_{\mathbb R} 2W \sinc(2 W \tau) \, x(t-\tau)\, d\tau.
    \label{eq:zt_relation}
\end{align}
Regular sampling gives $z_i=z(iT)$, $1\le i\le N$,  and $\zv=f(\cv)$ where $f(.)$ is based on \eqref{eq:zt_relation}. This mapping is invertible because the number of data-dependent sign changes per interval is at most one, see \cite[Lemma~1]{peleg2021information}. 

The schemes in \cite{peleg2021information} operate as follows. The best rates are achieved by Scheme D.
\begin{itemize}
    \item Scheme A has $c_i\in [(i-1)T,iT)$, and uses $d_i=iT$ for $i=1,\dots,N$, so $N_d=N$.
    \item Scheme B inverts the signal from Scheme A at times $iT$, so $N_d=0$, which eliminates every 2nd sign change. 
    \item Scheme D modifies Scheme B by requiring $c_i-c_{i-1}\ge T_g$ for some $T_g>0$. For example, \cite{peleg2021information} used $T_g=(0.2)T$. 
    \item Scheme C is different: it modulates Scheme A's signal by signs $s_i=\pm 1$ to send additional information. The mapping from $(c^n,s^n)$ to $x(t)$ is not invertible, so additional steps are required to compute achievable rates \cite{peleg2021information}.
\end{itemize}

%%%%%%%%%%%%%%%%%%%%%%%%%%%%%%%%%%% %%%%%%%%%%%%%%%%%%%%%%%%%%%%%%%%%%%
\section{Main Results}
\label{sec:main}

We simplify by setting $T=1$ in the following.

%%%%%%%%%%%%%%%%%%%%%%%%%%%%%%%%%%%
\subsection{Amplitude Modulation}
\label{subsec:AM}

We show that introducing amplitude modulation to \eqref{eq:schemes} does not increase $h(\Zm)$. Consider 
$N$ data-dependent amplitudes 
$$\av = (a_1,a_2,\dots,a_N)$$
at times $\cv$, and $N_d$ deterministic amplitudes 
$$\bv = (b_1,b_2,\dots,b_{N_d})$$
at times $\dv$. The amplitudes satisfy $0\le a_i\le 1$ and $0\le b_i\le 1$. Let $\av'$ be the amplitudes $(\av, \bv)$ ordered in the same way as $\cv'$. We write \eqref{eq:schemes} as
\begin{align*}
     x(t) & = a_1'u(t) + \sum_{i=1}^{N+N_d}
     (-1)^{i}\, (a_i'+a_{i+1}')\, u(t-c_i') \nonumber \\
     & \quad + (-1)^{N+N_d+1}\, a_{N+N_d+1}'\, u(t-N)
     %\label{eq:proof_level_01}
\end{align*}
where $a_{N+N_d+1}'$ may be chosen freely. We have the mapping $\zv=f(\cv, \av)$, and $f(.)$ remains invertible as long as the average number of sign changes per interval (of any signal difference) is at most one; see \cite[Lemma~1]{peleg2021information}.

The differential entropy is
\begin{align}
    h(\Zm) & = \int - p(\zv) \log p(\zv) \, d\zv \nonumber \\
    & =  \int p(\av) \left( \int - p(\zv|\av) \log p(\zv)\, d\zv \right) d\av .
    \label{eq:proof_level_03}
\end{align}
Consider fixed $\av$ and write $\cv(\zv)$ for the unique $\cv$ corresponding to $\zv$. We have
\begin{align}
    p(\zv|\av) = \left|\det \Jm(\cv(\zv)) \right|^{-1}  p_{\Cv|\Av}(\cv(\zv)| \av)
    \label{eq:proof_level_04}
\end{align}
with Jacobian $\Jm(\cv')=(\partial \zv / \partial \cv)_{\cv=\cv'}$ where the derivatives are
\begin{align}
    \frac{\partial z_i}{\partial c_j}
    & =  (-1)^{j+1} (a_{k(j)}'+a_{k(j)+1}') \sinc(i-c_j)
    \label{eq:proof_level_05}
\end{align}
for $1\le j\le N$, where $k(j)$ is the amplitude index corresponding to the $j$th time $c_j$, i.e., $a_{k(j)}'=a_j$. Inserting \eqref{eq:proof_level_05} into \eqref{eq:proof_level_04} gives
\begin{align}
    p(\zv|\av) =
    \frac{p(\cv(\zv)|\av)}{|\det \Sm_{\cv(\zv)} | \cdot \prod_i (A_{k(i)}'+A_{k(i)+1}')/2} 
    \label{eq:proof_level_06}
\end{align}
where $\Sm_{\cv}$ is a matrix with entries
\begin{align*}
    (\Sm_{\cv})_{i,j} = 2\sinc(i-c_j).
    %\label{eq:Sc-entries}
\end{align*}
From \eqref{eq:proof_level_06}, and using Bayes rule $p(\cv|\av)=p(\cv)P(\av|\cv)/P(\av)$ or similarly for continuous $\Av$, we have
\begin{align*}
    p(\zv) = \frac{p(\cv(\zv))}{|\det \Sm_{\cv(\zv)} |} \cdot
    \E{\left. \frac{1}{\prod_i (A_{k(i)}'+A_{k(i)+1}')/2} \right| \Cv=\cv(\zv)}
    %\label{eq:proof_level_07}
\end{align*}
where the expectation is over all $\av$ for which $\Cm=\cv(\zv)$ (this might involve one $\av$ only).
We further have
\begin{align*}
    d\zv = \left(|\det \Sm_{\cv(\zv)} | \cdot \prod\nolimits_i (A_{k(i)}'+A_{k(i)+1}')/2 \right) d\cv .
    %\label{eq:proof_level_08}
\end{align*}
Combining the above, the expression \eqref{eq:proof_level_03} becomes
\begin{align}
    & h(\Zm)
    = h(\Cv) + \int p(\cv) \log|\det \Sm_{\cv} |\, d\cv \nonumber \\
    & + \int - p(\cv) \log \E{\left. \frac{1}{ \prod_i (A_{k(i)}'+A_{k(i)+1}')/2} \right| \Cv =\cv} d\cv 
    \label{eq:proof_level_09}
\end{align}
and applying Jensen's inequality gives
\begin{align}
    &- \log \E{\left. \frac{1}{ \prod_i (A_{k(i)}'+A_{k(i)+1}')/2 } \right| \Cv =\cv} \nonumber\\
    &\qquad\le \sum\nolimits_i \E{\log \big((A_{k(i)}'+A_{k(i)+1}')/2\big) |\Cv =\cv} .
    \label{eq:proof_level_10}
\end{align}
Inserting \eqref{eq:proof_level_10} into \eqref{eq:proof_level_09}, and again applying Jensen's inequality, we have the bound
\begin{align}
    & h(\Zm)
    \le h(\Cv) + \int p(\cv) \log|\det \Sm_{\cv} |\, d\cv \nonumber \\
    & \qquad + \frac{N}{2} \log\left(\frac1N \sum_{i=1}^N \E{\left(\frac{A_{k(i)}'+A_{k(i)+1}'}{2}\right)^2}\right)
    \label{eq:proof_level_11}
\end{align}
with equality if the $A_i'$ are all the same constant. For example, for the peak constraint $A_i' \le 1$, we achieve the largest upper bound in \eqref{eq:proof_level_11} with $A_i'=1$, i.e., level-constrained signaling maximizes $h(\Zm)$.
Similarly, for the average power constraint $\sum_i \mathbb{E}[(A_i')^2]/N \le 1$, we achieve the largest upper bound in \eqref{eq:proof_level_11} with $A_i'=1$. Thus, level-constrained signaling maximizes $h(\Zm)$ under an average power constraint.

%%%%%%%%%%%%%%%%%%%%%%%%%%%%%%%%%%%
\subsection{Optimized PTS}
\label{subsec:pts}

The right-hand side of \eqref{eq:proof_level_11} with $A_i'=1$ for all $i$ gives
\begin{align}
    h(\Zm) = h(\Cv) + \E{\log|\det \Sm_{\Cm} |}.
    \label{eq:hZ-two-terms}
\end{align}
We wish to maximize $h(\Zm)$ over $p(\cv)$. For example, i.i.d. and uniformly-distributed $\Cv$ maximize $h(\Cv)$, while uniformly-spaced sampling times $c_j = j$ maximize $\E{\log|\det \Sm_{\Cm} |}$. However, neither choice optimizes $h(\Zm)$ in general.

To determine the optimal $p(\cv)$, rewrite \eqref{eq:hZ-two-terms} as
\begin{align}
    h(\Zm)
    & = \log\left(\int|\det \Sm_{\cv} |\, d\cv \right)-D(p(\cv)||p^*(\cv)) \label{eq:h_Z_and_D}
\end{align}
where the second term is an informational divergence with
\begin{align}
    p^*(\cv) = \frac{|\det \Sm_{\cv} |}{\int |\det \Sm_{\cv} |\, d\cv}. \label{eq:optimal_pts}
\end{align}
The optimal $p(\cv)$ is therefore $p^*(\cv)$. 

To compute \eqref{eq:optimal_pts}, let $\Mm$ be the Cauchy matrix with entries
\begin{align*}
    (\Mm)_{i,j} = \left(\frac{1}{i-c_j}\right)_{i,j}
\end{align*}
which has the determinant
\begin{align*}
    \det \Mm  = \frac{\prod_{j=2}^N (j-1)! \prod_{i<j} (c_i-c_j)}{\prod_{i,j} (i-c_j)}.
    %\label{eq:Cauchy-det}
\end{align*}
Recall that $c_i<c_j$ for $i<j$. We compute
\begin{align}
    & |\det \Sm_{\cv}| = \left| \det\left(2\frac{\sin(\pi(i-c_j))}{\pi (i-c_j)}\right)_{i,j} \right| \nonumber\\
    & = \left| \prod\nolimits_{j=1}^N \frac{2}{\pi} \sin(\pi c_j) \right| \cdot |\det \Mm| \nonumber \\
    & = \left( \prod\nolimits_{j=1}^N 2 (j-1)! \cdot \phi(c_j) \right) \prod_{1 \le i < j \le N} (c_j-c_i)
    \label{eq:Cauchy-matrix-expression}
\end{align}
where 
\begin{align*}
    \phi(x) = \frac{\sin(\pi x)/\pi}{\prod_{i=1}^N (i-x)}.
    %\label{eq:phi}
\end{align*}
Note that we discarded the absolute values in \eqref{eq:Cauchy-matrix-expression}. This step follows by applying L'Hôpital's rule at the integer values $x=k=1,\dots,N$ to obtain
\begin{align*}
    \phi(k) = \lim_{x \rightarrow k} \phi(x)
    = \frac{\cos(\pi k)}{-\prod_{i\ne k} (i-k)} 
    > 0.
\end{align*}
The signs of the numerator and denominator of $\phi(x)$ do not change for $x \in [k-1, k)$, so we have $\phi(x)\ge 0$ for $0\le x\le N$. Note also that the last product in \eqref{eq:Cauchy-matrix-expression} is the determinant of a Vandermonde matrix with $i,j$ entries $c_i^{j-1}$.

Next, define the functions
\begin{align*}
    \phi_{i}(x) = \phi(x) \cdot x^{i-1}
    %\label{eq:phi_i}
\end{align*}
and let $(\Phi)_{i,j}=\phi_{i}(c_j)$. Using \eqref{eq:Cauchy-matrix-expression}, we have
\begin{align}
    \int|\det \Sm_{\cv} | \, d\cv
    = \left( \prod\nolimits_{j=1}^N 2 (j-1)!\right) \int \det\Phi \, d\cv .
    \label{eq:integral-hz}
\end{align}
We use the general formulation in \cite[Eq.~(1.2)]{de1955some} to write
\begin{align}
    \idotsint\limits_{a \le c_1 < \cdots < c_N \le b} \det\Phi \, d\xv
    = \sqrt{\det \Am}
    \label{eq:deBruijn}
\end{align}
where $\Am$ is a particular skew-symmetric matrix, and $\sqrt{\det \Am}$ is the Pfaffian of $\Am$. For even $N$, we have
\begin{align}
    (\Am)_{i,j} = \int_a^b \int_a^b \phi_i(x)\,\phi_j(y)\,\sgn(y-x)\, dx\,dy
    \label{eq:A-entries}
\end{align}
for all $i,j$. For odd $N$, one instead considers the $(N+1)\times(N+1)$ matrix $\Am$ with the entries \eqref{eq:A-entries} for $1\le i, j\le N$ and
\begin{align}
    (\Am)_{i,N+1} = - (\Am)_{N+1,i}
    = \int_a^b \phi_i(x)\, dx
    \label{eq:A-entries-odd}
\end{align}
for $1\le i\le N$ and $(\Am)_{N+1,N+1}=0$. For example, numerical integration for $N=1,2,3,4,5$ gives the respective
\begin{align*}
    \gamma_1 \approx 0.0814, 0.1589, 0.2074, 0.2397, 0.2627.
\end{align*}

To aid computation for larger $N$, we prove the following Theorem in the Appendix. Define the modified Sine integrals
\begin{align*}
    \si_i(\alpha,\beta) &:= \int_\alpha^\beta
    \frac{\sin(x-i\pi)}{x-i\pi}\,dx 
    %\label{eq:sin-i}
    \\
    \si_{i,j}(\alpha,\beta) &:=  \int_\alpha^\beta \frac{\sin(x-i\pi)}{x-i\pi} \si_j(\alpha,x) \,dx .
    %\label{eq:sin-ij}
\end{align*}
Observe that $\si_i(\alpha,\beta)=\si(\beta-i\pi) - \si(\alpha-i \pi)$, where $\si(x)$ is the Sine integral. Also, integration by parts based on the derivative $d(\si_i(\alpha,x) \si_j(\alpha,x))/dx$ gives
\begin{align}
    \si_i(\alpha,\beta) \si_j(\alpha,\beta)
    = \si_{i,j}(\alpha,\beta) + \si_{j,i}(\alpha,\beta) .
    \label{eq:int-by-parts}
\end{align}

\begin{theorem}
    \label{thm:new-LB}
    Coding with $N$ sign changes in $N/(2W)$ seconds and the \gls{pts} density \eqref{eq:optimal_pts} gives
    \begin{align*}
        h(\Zm) = N \ln(2/\pi)
        + \ln(\det \Omegam )/2
        %\label{eq:optimum_gamma}
    \end{align*}
    where, for even $N$, $\Omegam$ is a $N \times N$ skew-symmetric matrix with 
    \begin{align}
        (\Omegam)_{i,j} & =  \si_{j,i}(0,N\pi) - \si_{i,j}(0,N\pi)
        \label{eq:omega-entries-even}
    \end{align}
    for all $i,j$. For odd $N$, $\Omegam$ is an $(N+1)\times(N+1)$ skew-symmetric matrix with entries \eqref{eq:omega-entries-even} for $1\le i,j\le N$ and
    \begin{align}
        (\Omegam)_{i,N+1} 
        = -(\Omegam)_{N+1,i}
        =  \si_i(0,N\pi)
        \label{eq:omega-entries-odd}
    \end{align}
    for $1\le i\le N$ and $(\Omegam)_{N+1,N+1}=0$.
\end{theorem}

\begin{figure}[!t]
    \centering
    \includegraphics[width=0.485\textwidth]{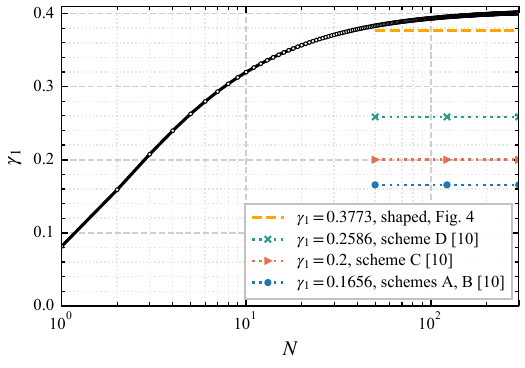}
    \caption{Power factor $\gamma_1$ vs. block length $N$.}
    \label{fig:gamma_1}
\end{figure}

Theorem~\ref{thm:new-LB} replaces $\Am$ with $\Omegam$, and we find that $(\Omegam)_{i,j}\approx 2 \pi \si(\pi (j-i))$ for large $N$. Also, the identity \eqref{eq:int-by-parts} and skew-symmetry reduce the number of integrals needed to compute $\Omegam$. Fig. \ref{fig:gamma_1} plots $\gamma_1$ for $N\le 300$. Observe that $\gamma_1$ increases monotonically with $N$; we computed $\gamma_1 \approx 0.4039$ for $N=1000$. Asymptotic considerations via Szeg\"o's theorem for Toeplitz matrices suggest that $\lim_{N\rightarrow\infty} \gamma_1 = 4/\pi^2 \approx 0.4053$.

\begin{figure}[!t]
    \centering
    \includegraphics[width=0.485\textwidth]{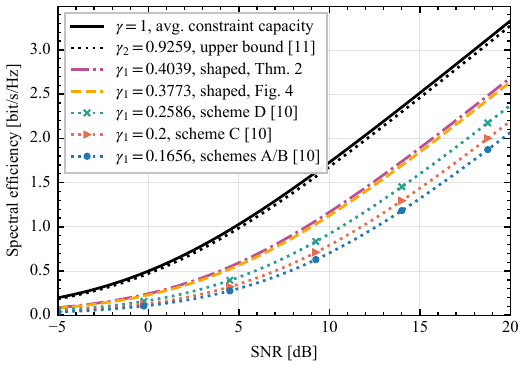}
    \caption{Rate bound comparisons.}
    \label{fig:rate_comp}
\end{figure}

The corresponding rate bounds \eqref{eq:EPI2} are shown in Fig. \ref{fig:rate_comp}. The three lower dashed curves are the bounds for the four schemes in \cite{peleg2021information}. The dash-dotted curve is our improved lower bound that gains 1.94~dB over the curve of Scheme D. The gain of the asymptotic $\gamma_1 = 4/\pi^2$ is almost the same at 1.95~dB.

%%%%%%%%%%%%%%%%%%%%%%%%%%%%%%%%%%%%%%%%%%%
%%%%%%%%%%%%%%%%%%%%%%%%%%%%%%%%%%%%%%%%%%%
\subsection{Sequential PTS}
\label{subsec:pts-sequential}

One can implement \gls{pts} by sequentially encoding with the conditional densities $p^*(c_i | c^{i-1})$. We simplify by optimizing a unit memory density $p(c_i | c_{i-1})$ as follows.
\begin{itemize}
    \item Choose the support of $p(c_i | c_{i-1})$ as $[c_{i-1}, i+1)$ to position $c_i$ near sampling time $i$. 
    \item Split $[c_{i-1}, i+1)$ into $K$ sub-intervals of equal length $\Delta=(i+1-c_{i-1})/K$, and assign each sub-interval a probability $P(k)$, $1\le k\le K$. 
    \item Choose $p(c_i | c_{i-1})$ by linearly interpolating between the $K+2$ points
    \begin{align*}
        (c_{i-1},0), (t_1,P(1)), \dots, (t_K,P(K)), (i+1,0)
    \end{align*}
    where $t_k=c_{i-1}+(k-1/2)\Delta$,
    and normalize to a density. The entropy $h(C_i|C_{i-1}=c_{i-1})$ is then
    \begin{align}
        \sum_{k=0}^K \int_{t_k}^{t_{k+1}} -(a_k+b_k t) \log(a_k+b_k t)\, dt
        \label{eq:entropy-interpolation}
    \end{align}
    where $t_0=c_{i-1}$, $t_{K+1}=i+1$, and $p(c_i|c_{i-1})=a_k + b_k c_i$ in the $k$th sub-interval with coefficients $a_k,b_k$.
\end{itemize}

We used coordinate descent to optimize $(P(1),\dots,P(K))$. Each $P(k)$ was initialized using a uniform distribution on $\mathcal{P}_L=\{0, \frac{1}{L}, \frac{2}{L},\dots,1\}$, and then normalized by $\sum_k P(k)$. We next successively optimized the $P(k)$, $k=1,\dots,K$, while keeping the other $P(m)$ fixed. We computed \eqref{eq:hZ-two-terms} by Monte Carlo integration for each choice of $P(k)$ from $\mathcal{P}_L$ by normalizing by $\sum_k P(k)$, interpolating to obtain $p(c_i|c_{i-1})$, computing \eqref{eq:entropy-interpolation} for each $c_i$ chosen randomly from $p(c_i|c_{i-1})$, and averaging across $i=1,\dots,N$. This process was repeated until the $P(k)$ and $p(c_i|c_{i-1})$ hardly changed.

\begin{figure}[!t]
    \centering
    \includegraphics[width=0.485\textwidth]{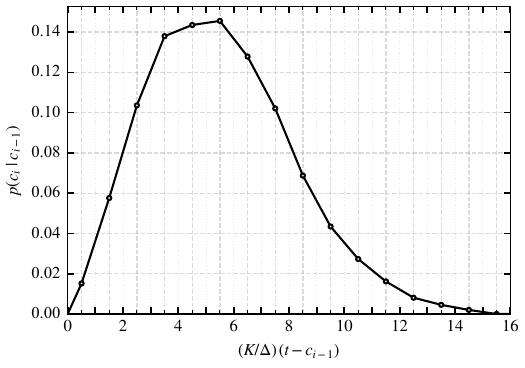}
    \caption{Optimized \gls{pts} density for $K=16$ and $L=100$. The sub-interval lengths are $\Delta=(i+1-c_{i-1})/K$.}
    \label{fig:tradeoff_numerical}
\end{figure}

Fig. \ref{fig:tradeoff_numerical} shows the result for $K = 16$ and $L=100$. This sequential \gls{pts} scheme achieves $\gamma_1 \approx 0.3773$, improving $\gamma_1 \approx 0.2586$ by $1.64$ dB. We remark that the support of $p(c_i|c_{i-1})$ is time-varying. Two simple extensions are to expand the support of $p(c_i|c_{i-1})$ beyond time $i+1$, and to increase the memory.

%%%%%%%%%%%%%%%%%%%%%%%%%%%%%%%%%%%
%%%%%%%%%%%%%%%%%%%%%%%%%%%%%%%%%%%
\section{Conclusion}
\label{sec:conclusion}

We studied level-constrained signaling over bandlimited \gls{awgn} channels and showed that \gls{pts} improves the best existing lower bound on capacity by at least 1.94 dB at high $\SNR$. A simpler sequential scheme gains 1.64 dB at high $\SNR$. We considered signals having one sign change per Nyquist-rate sample on average. One can expect there are better schemes for a larger number of sign changes.

%%%%%%%%%%%%%%%%%%%%%%%%%%%%%%%%%%% %%%%%%%%%%%%%%%%%%%%%%%%%%%%%%%%%%%
%\clearpage
\appendices
\section*{Appendix}
We prove Theorem \ref{thm:new-LB}. Define the integral
\begin{align*}
    \Phi_i(\alpha,\beta)
    & := \int_\alpha^\beta \phi_i(x) \, dx \nonumber \\
    & = 
    \int_\alpha^\beta 
    \underbrace{\left(\frac{x^{i-1}}{\prod_{j=1}^N (j-x)}\right)}_{\displaystyle R(x)} \, \frac{\sin(\pi x)}{\pi} \, dx.
\end{align*}
Using residue theory, for $1\le i\le N$, we can expand
\begin{align}
    R(x) = \sum_{k=1}^N (-1)^k \, \frac{k^i}{N!} \,
    \binom{N}{k} \,\frac{1}{x-k}
    \label{eq:rational}
\end{align}
and thus obtain
\begin{align}
    \Phi_{i}(\alpha,\beta) 
    & = \sum_{k=1}^N \frac{k^i}{\pi N!}
    \binom{N}{k} \, \underbrace{\int_\alpha^\beta 
    \frac{\sin(\pi(x-k))}{x-k}\, dx}_{\displaystyle \si_k(\alpha \pi, \beta \pi)} .
    \label{eq:integral1}
\end{align}

The double integral \eqref{eq:A-entries} can be written as
\begin{align*}
    (\Am)_{i,j} 
    &= \underbrace{\int_a^b \phi_j(y)\,
    \Phi_i(a,y)\,dy}_{\displaystyle B_{j,i}} -  \underbrace{\int_a^b \phi_i(x)\,
    \Phi_j(a,x)\,dx}_{\displaystyle B_{i,j}} .
    %\label{eq:Pfaffian_entries}
\end{align*}
Using \eqref{eq:rational} we can write
\begin{align*}
    B_{i,j} 
    &= \sum_{k=1}^N \frac{k^i}{\pi N!}  \binom{N}{k} \int_a^b \frac{\sin(\pi (x-k))}{x-k} \, 
    \Phi_{j}(a,x) \,dx \nonumber \\
    & = \sum_{k=1}^N\sum_{m=1}^N 
    \frac{k^i m^j}{(\pi N!)^2} \, 
    \binom{N}{k} \binom{N}{m} \, \si_{k,m}(a\pi, b\pi) .
    % \label{eq:integral2}
\end{align*}

Now use $a=0$ and $b=N$ to write (see \eqref{eq:omega-entries-even})
\begin{align*}
    (\Am)_{i,j}
    & = \sum_{k=1}^N\sum_{m=1}^N 
    \frac{k^i m^j}{(\pi N!)^2} \, 
    \binom{N}{k} \binom{N}{m} \,
    (\Omegam)_{k,m} .
    %\label{eq:Aij}
\end{align*}
For even $N$, we therefore have
\begin{align}
    \Am = \frac{1}{(\pi N!)^2} \Vm^T\, \Dm\, \Omegam\, \Dm\, \Vm
    \label{eq:A}
\end{align}
where $\Dm$ is diagonal with entries $(\Dm)_{k,k}=k \binom{N}{k}$ and $\Vm$ is Vandermonde with entries $(\Vm)_{k,i}=k^{i-1}$. We compute
\begin{align}
    \det\Am
    & = \frac{1}{(\pi N!)^{2N}} \big( \det\Vm \cdot \det\Dm \big)^2  \det\Omegam \nonumber\\
    & = \left(\prod_{1\le i < j \le N} (j-i) \cdot \prod\nolimits_{j=1}^N \frac{j}{\pi N!} \binom{N}{j} \right)^2 \det\Omegam
    \nonumber\\
    & = \left(\prod\nolimits_{j=1}^N \frac{1}{\pi\, (j-1)!} \right)^2 \det\Omegam .
    \label{eq:detA-even}
\end{align}
Using \eqref{eq:h_Z_and_D}, \eqref{eq:integral-hz}, and \eqref{eq:deBruijn}, we obtain
\begin{align}
    h(\Zm) 
    & = \log \left( \left( 
    \prod\nolimits_{j=1}^N 2 (j-1)! \right) \sqrt{\det\Am} \right)
    \nonumber \\
    & = N\ln\left(\frac{2}{\pi}\right) + \frac{1}{2}\ln( \det\Omegam ) .
    \label{eq:hZ-even}
\end{align}

For odd $N$, we must consider the augmented matrix 
\begin{align}
    \Am & =
    \begin{bmatrix}
        \Am_N & \av_{N+1} \\ -\av_{N+1}^T & 0
    \end{bmatrix} 
    \label{eq:A-odd}
\end{align}
where $\Am_N$ is the $N\times N$ matrix with entries \eqref{eq:A-entries} and 
\begin{align}
    \av_{N+1} & = [\Phi_1(0,N)\;\dots\;\Phi_N(0,N)]^T.
    \label{eq:p}
\end{align}
Using \eqref{eq:A}, we have
\begin{align}
    \Am_N = \frac{1}{(\pi N!)^2} \Vm^T\, \Dm\, \Omegam_N\, \Dm\, \Vm 
    \label{eq:AN}
\end{align}
where $\Am_N$ and $\Omegam_N$ both have rank $N-1$, i.e., they each have one zero eigenvalue. Note that $\av_{N+1}$ must lie outside the column space of $\Am_N$ for $\Am$ to have non-zero determinant.

We now use \eqref{eq:AN} to write
\begin{align}
    \Am & = \frac{1}{(\pi N!)^2}
    \begin{bmatrix}
        \Vm^T \Dm & {\bf 0} \\ {\bf 0}^T & c
    \end{bmatrix}
    \begin{bmatrix}
        \Omegam_N & \omegav \\ -\omegav^T & 0
    \end{bmatrix}
    \begin{bmatrix}
        \Dm \Vm & {\bf 0} \\ {\bf 0}^T & c
    \end{bmatrix} \nonumber \\
    & = \begin{bmatrix}
        \Am_N 
        & \frac{c}{(\pi N!)^2} \Vm^T \Dm\, \omegav \\
        - \frac{c}{(\pi N!)^2} \omegav^T \Dm \Vm & 0
    \end{bmatrix}
    \label{eq:A-odd2}
\end{align}
for some $c\ne 0$. The identities \eqref{eq:A-odd}, \eqref{eq:p}, and \eqref{eq:A-odd2} give
\begin{align*}
    \Phi_i(0,N)
    & = \frac{c}{(\pi N!)^2}
        \big( \Vm^T \Dm\, \omegav \big)_k
    \nonumber \\
    & = \frac{c}{(\pi N!)^2} \sum_{k=1}^N k^i \binom{N}{k}\, \omega_k.
    %\label{eq:omega_define}
\end{align*}
Thus, using \eqref{eq:integral1} and choosing $c=(\pi N!)$, we have
\begin{align*}
    \omegav & = [\si_1(0,N\pi)\;\dots\;\si_N(0,N\pi)]^T.
\end{align*}
The determinant of \eqref{eq:A-odd2} is again \eqref{eq:detA-even}, and $h(\Zm)$ is again \eqref{eq:hZ-even}, except that
\begin{align*}
    \Omegam = \begin{bmatrix}
        \Omegam_N & \omegav \\ -\omegav^T & 0
    \end{bmatrix}.
    %\label{eq:omegam-odd}
\end{align*}

We may obtain more insight for odd $N$. Observe that
\begin{align*}
    \Omegam_N \adj(\Omegam_N)
    = (\det\Omegam_N)\, \Id_N = 0
    %\label{eq:adj_01}
\end{align*}
where $\adj(\Omegam_N)$ is the adjugate of $\Omegam_N$. Thus, the columns of $\adj(\Omegam_N)$ are in the null space of $\Omegam_N$. But the null space has dimension one, implying that $\adj(\Omegam_N)=\uv\,\uv^T$ for some $\uv$. We further have (see \cite[p.~26]{horn2012matrix})
\begin{align*}
    \det\Omegam
    = \omegav^T (\uv\,\uv^T)\, \omegav 
    = (\uv^T \omegav)^2 .
    %\label{eq:det_Omega_odd}
\end{align*}

%%%%%%%%%%%%%%%%%%%%%%%%%%%%%%%%%%% %%%%%%%%%%%%%%%%%%%%%%%%%%%%%%%%%%%
\section*{Acknowledgment}
This work was supported by the German Research Foundation (DFG) under Project KR 3517/13-1 and SH 1937/1-1.

%%%%%%%%%%%%%%%%%%%%%%%%%%%%%%%%%%%
%%%%%%%%%%%%%%%%%%%%%%%%%%%%%%%%%%%
\small
\bibliographystyle{IEEEtran}
\bibliography{LC-Ref}

\end{document}